\documentclass[preprint,5p,twocolumn,authoryear]{elsarticle}

\usepackage{amsmath,amssymb,amsfonts,dsfont,amsthm}
\usepackage{cmap}
\usepackage{hyperxmp}
\usepackage[pdfdisplaydoctitle=true,
      colorlinks=true,
      urlcolor=blue,
      citecolor=blue,
      linkcolor=blue,
      pdfstartview=FitH,
      pdfpagemode=None,
      bookmarksnumbered=true]{hyperref} 

\newtheorem{thm}{Theorem}[section]

\theoremstyle{definition}

\theoremstyle{remark}
\newtheorem{rem}[thm]{Remark}

\numberwithin{equation}{section}

\newcommand{\nn}{\pmb{n}}                   
\newcommand{\ww}{\pmb{w}}                   
\newcommand{\xx}{\pmb{x}}                   
\newcommand{\bnu}{\pmb{\nu}}                

\newcommand{\Idd}{\mathbf{1}}               
\newcommand{\rh}{\mathrm{h}}                

\newcommand{\ba}{\mathbf{a}}
\newcommand{\bb}{\mathbf{b}}
\newcommand{\bc}{\mathbf{c}}
\newcommand{\bd}{\mathbf{d}}
\newcommand{\bh}{\mathbf{h}}

\newcommand{\bepsilon}{\pmb \epsilon}
\newcommand{\bsigma}{\pmb \sigma}
\newcommand{\bomega}{\pmb \omega}
\newcommand{\bPhi}{\pmb \Phi}
\newcommand{\bphi}{\pmb \phi}

\newcommand{\bH}{\mathbf{H}}                
\newcommand{\bT}{\mathbf{T}}                
\newcommand{\bD}{\mathbf{D}}                
\newcommand{\bF}{\mathbf{F}}                
\newcommand{\bOmega}{\pmb{\Omega}}          

\DeclareMathOperator{\tr}{tr}

\newcommand{\norm}[1]{\left\Vert#1\right\Vert}  
\newcommand{\set}[1]{\left\{#1\right\}}         

\newcommand{\tq}[1]{{{\mathbf{#1}}}}
\newcommand{\td}[1]{{\mathbf{#1}}}
\renewcommand{\vec}{\pmb}

\newcommand{\MES}{\mathcal{R}}

\def\scal{\mbox{\,\scriptsize$\mathsf\bullet$\;}}
\newcommand{\otimesbar}{\; \underline{\overline{\otimes}} \;}

\newcommand{\beq}{\begin{equation}}
\newcommand{\eeq}{\end{equation}}

\journal{European Journal of Mechanics A/Solids (accepted, https://doi.org/10.1016/j.euromechsol.2017.11.014)}

\hypersetup{
  pdfauthor={R. Desmorat, B. Desmorat, M. Olive and B. Kolev}, 
  pdftitle={Micromechanics based 3D framework with second-order damage tensors}, 
  pdfsubject={}, 
  pdfkeywords={crack density \sep damage tensor \sep harmonic decomposition}, 
  pdflang=en, 
  }
\begin{document}

\begin{frontmatter}



  \title{Micromechanics based framework \\
  with second-order damage tensors}


  \author[label1]{R. Desmorat}
  \ead{desmorat@lmt.ens-cachan.fr}
  \author[label2,label3]{B. Desmorat}
  \ead{boris.desmorat@upmc.fr}
  \author[label1]{M. Olive}
  \ead{marc.olive@math.cnrs.fr}
  \author[label4]{B. Kolev}
  \ead{boris.kolev@math.cnrs.fr}

  \address[label1]{LMT-Cachan (ENS Cachan, CNRS, Universit\'{e} Paris Saclay), F-94235 Cachan Cedex, France}
  \address[label2]{Sorbonne Universit\'e, UMPC Univ Paris 06, CNRS, UMR 7190, Institut d'Alembert,F-75252 Paris Cedex 05, France}
  \address[label3]{Universit\'e Paris SUD 11, Orsay, France}
  \address[label4]{Aix Marseille Universit\'{e}, CNRS, Centrale Marseille, I2M, UMR 7373, 13453 Marseille, France}

  \begin{abstract}
    The harmonic product of tensors---leading to the concept of harmonic factorization---has been defined in a previous work (Olive et al, 2017). In the practical case of 3D crack density measurements on thin {or thick} walled structures, this mathematical tool allows us to factorize the harmonic (irreducible) part of the fourth-order damage tensor as an harmonic square: an exact harmonic square in 2D, an harmonic square over the set of so-called\emph{ mechanically accessible directions for measurements} in the 3D case. The corresponding micro-mechanics framework based on second---instead of fourth---order damage tensors is derived. {An illustrating example is provided showing how the proposed framework allows for the modeling of the so-called hydrostatic sensitivity up to high damage levels.}
  \end{abstract}


\end{frontmatter}

\section{Introduction}
\label{sec:introduction}

The damage anisotropy encountered in quasi-brittle materials is induced by the loading direction and multiaxiality. From a micro-mechanics point of view, it is the consequence of an oriented microcracking pattern. From the Continuum Damage Mechanics (CDM) point of view, the anisotropic damage state is represented by a tensorial thermodynamics variable, either an eight-order tensor \citep{Cha1978,Cha1979}, a fourth-order damage tensor $\tq D$~\citep{Cha1978,Cha1979,LO1980,Cha1984,LC1985,ABM1986,Ju1989,Kac1993,ZC1998,CW2010,DK2016} or a symmetric second-order damage tensor $\bd$~\citep{VK1971,MO1978,CS1982,Lad1983,Mur1988}.

There exist many second-order anisotropic damage frameworks~\citep{Mur1988,KV1990,RBM1992,PT1996,HD1998,SC1998,LDS2000,CRW2001,MS2001,MESR2002,Bru2003,LD2005,DGR2007,BGL2007,DO2006,Des2016},
as their unification into a single model is partial~\citep{Lad1983,Lad1995}.  {A link with the theory of second order fabric tensors has been made in \citep{ZC1995,VK2006}.
From a theoretical point of view~\citep{LO1980,Ona1984}, second order damage frameworks are usually seen to be restrictive compared to the fourth-order tensorial one.} Nevertheless, the interpretation of a damage variable being simpler when a second-order tensor is considered (the three principal values $d_{i}$ of $\bd$ naturally correspond to 3 orthogonal families of microcracks), less damage parameters are introduced and the second-order frameworks have been widely used for either ductile or quasi-brittle materials.

The recent analysis of 2D cracked media with both open and closed microcraks has shown that the so-called irreducible (harmonic) part $\bH_{2D}$ of the damage tensor can be decomposed by means of a second-order damage tensor~\citep{DD2016}. More precisely, the standard second-order crack density tensor of~\cite{VK1971} still represents the open cracks contribution when a novel (deviatoric) second-order damage tensor represents the closed---sliding---cracks (previously represented by a fourth-order tensor, \cite{ABM1986,Kac1993}). This can be achieved using Verchery's polar decomposition of 2D fourth-order tensors~\citep{Ver1979,Van2005}, which includes both~\citep{DD2015}:
\begin{itemize}
  \item[--]   the harmonic decomposition of considered tensor;
  \item[--]   the harmonic factorization of its fourth-order irreducible (harmonic) part $\bH_{2D}$, by means of a deviatoric second-order tensor $\bh_{2D}$:
        \begin{equation*}
          \bH_{2D}=\bh_{2D}\ast \bh_{2D}.
        \end{equation*}
\end{itemize}
The harmonic product between harmonic tensors, written as $\bh_1 \ast \bh_2$, is defined as the projection of the (totally) symmetric tensor product $\bh_1\odot \bh_2$ onto the space of harmonic tensors (see Sections~\ref{sec:Sylvester-theorem} and~\ref{subsec:2D-case}).

The question arises then as how to extend these results in 3D ? We know from~\citep{OKDD2016} that any 3D harmonic fourth-order tensor can be factorized into
\begin{equation*}
  \bH = \bh_1 \ast \bh_2,
\end{equation*}
\textit{i.e.}, represented by two (deviatoric) second-order tensors $\bh_1, \bh_2$. However, the factorization is far from being unique.

To overpass these difficulties, we point out that triaxial mechanical testing is of high complexity, both from the experimental set-up needed (a triaxial machine) and from the difficulty to measure mechanical properties in different space directions~\citep{Cal1997,CM1999}.
We propose, here, to restrict ourselves to a simpler, but still sufficiently general, situation:
{
 the case of measurements of a 3D crack density function
on structures.  Well-known cases are the thin walled structures, such as plates, tubes and shells for which the thinner direction is the normal $\bnu$. But the present work also applies to 3D thick structures (as the cube of Section \ref{S:ShearBulk}), as long as an out-of-plane normal can be locally defined.
}

Instead of considering the representation of crack density in any direction $\nn$, we shall {then} consider, in Section~\ref{sec:3D-crack-density}, its representation to a restricted set of directions
\begin{equation*}
  \MES(\bnu) := \set{\pmb{\tau};\; \norm{\pmb{\tau}}=1 \; \textrm{and} \;  \pmb{\tau}\cdot\bnu=0} \cup \set{\bnu},
\end{equation*}
\textit{i.e.}, the \emph{in-plane directions} $\pmb{\tau}$, orthogonal to $\bnu$, and the \emph{out-of-plane direction} $\bnu$, normal to the structure itself. In the present work, we consider these directions as the \emph{mechanically accessible directions for measurements}.

After recalling the required mathematical tools (harmonic decomposition, harmonic product and Sylvester's theorem), we revisit the link between crack density function and the tensorial nature of the damage variables. This will allow us to derive a general micro-mechanics based 3D framework with second---instead of fourth---order damage tensors.

\subsection*{Definitions}

We denote by $\bT^\mathrm{s}$ the totally symmetric part of a possibly non symmetric tensor $\bT$. More precisely
\begin{equation*}
  \bT^\mathrm{s}(\xx_{1}, \dotsc , \xx_{n}) := \frac{1}{n!} \sum_{\sigma \in \mathfrak{S}_{n}} \bT(\xx_{\sigma(1)}, \dotsc , \xx_{\sigma(n)}),
\end{equation*}
where $\mathfrak{S}_{n}$ is the permutation group on the indices $\set{1,\dotsc , n}$. The \emph{symmetric tensor product} of two tensors $\bT_{1}$ and $\bT_{2}$, of respective orders $n_1$ and $n_2$, is the symmetrization of $\bT_{1} \otimes \bT_{2}$, defining a totally symmetric tensor of order $n=n_1+n_2$:
\begin{equation*}
  \bT_{1} \odot \bT_{2} := ( \bT_{1} \otimes \bT_{2} )^\mathrm{s}.
\end{equation*}

Contracting two indices $i,j$ of a tensor $\bT$ of order $n$ defines a new tensor of order $n-2$ denoted as $\tr_{ij} \bT$. For a totally symmetric tensor $\bT$, this operation does not depend on a particular choice of the pair $i,j$. Thus, we can refer to this contraction just as the \emph{trace} of $\bT$ and we will denote it as $\tr \bT$. It is a totally symmetric tensor of order $n-2$. Iterating the process, we define
\begin{equation*}
  \tr^{k} \bT = \tr(\tr(\dotsb (\tr \bT))),
\end{equation*}
which is a totally symmetric tensor of order $n-2k$.

In 3D, a totally symmetric fourth-order tensor $\bT$ has no more than 15 independent components, instead of 21 for a triclinic elasticity tensor (\textit{i.e.} a tensor $\tq{E}$ having minor symmetry $E_{ijkl}=E_{jikl}=E_{ijlk}$ and major symmetry $E_{ijkl}=E_{klij}$). Totally symmetric elasticity tensors were called \emph{rari-constant} in the nineteenth century~\citep{Nav1827,Cau1828a,Cau1828b,Poi1829,Lov1905,VD2016}.

\section{Harmonic decomposition}
\label{sec:harmonic-decomposition}

The harmonic decomposition of tensors~\citep{Sch1954,Spe1970}, introduced in geophysics by~\cite{Bac1970}, has been popularized by~\cite{LO1980} and~\cite{Ona1984} when deriving fourth-order damage tensor and later by~\cite{FV1996} when classifying elasticity symmetries.

\subsection{Harmonic tensors and corresponding polynomials}
\label{subsec:harmonic-tensors-versus-polynomials}

An harmonic tensor is a traceless, totally symmetric tensor, \textit{i.e.}
\begin{equation*}
  \bH=\bH^\mathrm{s}, \quad \text{and} \quad \tr \tq H=0.
\end{equation*}
To every totally symmetric tensor $\bH$ of order $n$, with components $H_{i_{1}i_{2}\dotsm i_{n}}$, corresponds a unique homogenous polynomial (and conversely). More precisely,
\begin{equation*}
  \rh (\vec x) := \bH(\vec x, \vec x, \dots, \vec x)=H_{i_{1}i_{2}\dotsm i_{n}} x_{i_{1}}x_{i_{2}}\dotsm x_{i_{n}}
\end{equation*}
is a homogeneous polynomial
\begin{equation*}
  \rh (\vec x) = \rh(x_1, x_2, x_3),
\end{equation*}
of degree $n$ in the spacial coordinates $x_1, x_2, x_3$. It is harmonic since $\nabla^2 \rh=0$, due to the traceless property $\tr \bH=0$.

\subsection{Harmonic decomposition of a symmetric tensor}
\label{subsec:totally-symmetric-tensors}

Any totally symmetric tensor $\bT$ of order $n$ can be decomposed uniquely as
\begin{equation}\label{eq:tensorial-symmetric-harmonic-decomposition}
  \bT = \bH_{0} + \Idd \odot \bH_{1} + \dotsb + \Idd^{\odot r-1}\odot \bH_{r-1}+ \Idd^{\odot r} \odot\bH_{r}
\end{equation}
where $r=[n/2]$ is the integer part of $n/2$, $\bH_{k}$ is an harmonic tensor of degree $n-2k$ and $\Idd^{\odot k}=\Idd \odot \dotsb \odot \Idd$ means the symmetrized tensorial product of $k$ copies of the (second-order) identity tensor.
For $n$ even ($n=2r$), one has:
\begin{equation}\label{eq:Hr}
  \bH_{r} = \bH_{\frac{n}{2}} = {\frac{1}{n+1}}\tr^\frac{n}{2} \bT,
\end{equation}
where $\bH_{r}=H_r$ is a scalar in that case. Moreover, $\bH_{r-1}, \dotsc ,\bH_{0}$ are obtained inductively~\citep{OKDD2016} as follows:
\begin{equation}\label{eq:Hk}
  \begin{split}
    \bH_{k}= \mu(k,n)
    \tr^{k}\Big[ \bT - \sum_{j=k+1}^{r} \Idd^{\odot j}\odot \bH_{j} \Big]
  \end{split}
\end{equation}
where $ \mu(k,n)=\displaystyle \frac{(2n-4k+1)!(n-k)!n!}{(2n-2k+1)!k!(n-2k)!(n-2k)!}$.

\begin{rem}
  It is worth emphasizing the fact that this harmonic decomposition is just a generalization to \emph{higher order symmetric tensors} of the well-known decomposition of a \emph{symmetric second-order} tensor into its \emph{deviatoric/spheric} parts:
  \begin{equation*}
    \bd = \bd' + \frac{1}{3}(\tr \bd) \Idd.
  \end{equation*}
\end{rem}

Decomposition~\eqref{eq:tensorial-symmetric-harmonic-decomposition} is an orthogonal decomposition (relative to the natural Euclidean product on the space of symmetric tensors). The projection $\bH_{0}$ onto the space of highest order harmonic tensors (same order as $\bT$) will be called the \emph{harmonic part} of $\bT$ and denoted by $(\bT)_{0}$:
\begin{equation}\label{eq:T0}
  (\bT)_{0} := \bH_{0} = \bT - \Idd \odot \bH_{1} - \dotsb  - \Idd^{\odot r} H_r.
\end{equation}

\subsection{Harmonic decomposition of the elasticity tensor}
\label{subsec:elasticity-tensor}

The harmonic decomposition of an elasticity tensor $\tq E$, a fourth-order tensor having both minor and major symmetries ($E_{ijkl}=E_{jikl}=E_{ijlk}$ and $E_{ijkl}=E_{klij}$), was first obtained by~\cite{Bac1970}, as:
\begin{equation}\label{eq:HarmE}
  \tq{E} = \alpha \, \Idd \otimes_{(4)} \Idd + \beta \, \Idd \otimes_{(2,2)}\! \Idd
  + \Idd \otimes_{(4)} \td a' + \Idd \otimes_{(2,2)} \! \td b' + \bH
\end{equation}
where $(\cdot)'=(\cdot)-\frac{1}{3}\tr(\cdot)\, \Idd$ denotes the deviatoric part of a second-order tensor.

In formula (\ref{eq:HarmE}), the Young-symmetrized tensor products $\otimes_{(4)}$ and $\otimes_{(2,2)}$, between two symmetric second-order tensors $\td y, \td z$, are defined as follows:
\begin{equation}\label{eq:product422}
  \begin{cases}
    \td y \otimes_{(4)} \td z = \frac{1}{6} \big( \td y \otimes \td z +  \td z \otimes \td y
    + 2\, \td y \otimesbar \td z + 2\, \td z \otimesbar \td y \big),                             \\
    \td y \otimes_{(2,2)}\!\td z  = \frac{1}{3} \big( \td y \otimes \td z +  \td z \otimes \td y
    - \td y \otimesbar \td z - \td z \otimesbar \td y \big),
  \end{cases}
\end{equation}
where $(\td y \otimesbar \td z)_{ijkl}=\frac{1}{2} (y_{ik}z_{jl} + y_{il}z_{jk})$ so that $\otimes_{(4)}$ is the same as the totally symmetric tensor product $\odot$:
\begin{equation*}
  \td y \otimes_{(4)} \td z =\td y \odot \td z.
\end{equation*}

In the harmonic decomposition~\eqref{eq:HarmE}, $\bH$ is a fourth-order harmonic tensor, $\alpha,\beta$ are scalars, and $\td{a}',\td{b}'$ are second-order harmonic tensors (deviators) related to the \emph{dilatation tensor} $\td {di}=\tr_{12} \tq E$ and the \emph{Voigt tensor} $\td {vo}=\tr_{13}\tq E$ by the formulas:
\begin{equation}\label{eq:alpha_beta}
  \alpha = \frac{1}{15}\left( \tr \td {di} + 2 \tr \td {vo}\right),
  \quad
  \beta = \frac{1}{6}\left( \tr \td {di} - \tr \td {vo}\right),
\end{equation}
and
\begin{equation}\label{eq:a_b}
  \td a'=\frac{2}{7} \left( \td {di}'+2 \td {vo} ' \right),
  \quad
  \td b'=2 \left( \td {di}'- \td {vo} ' \right).
\end{equation}

The harmonic part of $\tq E$ is defined as:
\begin{equation}\label{eq:E0}
  (\tq E)_{0} := \bH = \tq{E} - \Idd \otimes_{(4)} \ba -\Idd \otimes_{(2,2)}\! \bb
\end{equation}
or similarly as:
\begin{equation*}
  (\tq E)_{0} := \tq{E} - \Idd \odot \td a - \frac{1}{3} \left( \Idd \otimes \bb + \bb \otimes \Idd - \Idd \otimesbar \bb - \bb \otimesbar \Idd \right),
\end{equation*}
where $\ba= \ba' + \alpha \Idd $ and $\bb= \bb' + \beta \Idd$. The scalars $\alpha, \beta$ and the second-order deviators $\ba', \bb'$ are given by~\eqref{eq:alpha_beta} and~\eqref{eq:a_b}.

\section{The harmonic product and Sylvester's theorem}
\label{sec:Sylvester-theorem}

The harmonic product of two harmonic tensors of order $n_1$ and $n_2$, defining an harmonic tensor of order $n=n_1+n_2$, has been introduced in~\citep{OKDD2016} as the harmonic part of the symmetric tensor product:
\begin{equation*}
  \bH_1 \ast \bH_2:= \left( \bH_1 \odot \bH_2 \right)_{0}.
\end{equation*}
Note that this product is \emph{associative}:
\begin{equation*}
  \bH_1 \ast (\bH_2\ast \bH_3) = (\bH_1 \ast \bH_2) \ast \bH_3,
\end{equation*}
and \emph{commutative}:
\begin{equation*}
  \bH_1 \ast \bH_2=\bH_2 \ast \bH_1.
\end{equation*}

For two vectors $\vec w_1, \vec w_2$, we have
\begin{equation*}
  \begin{split}
    \ww_1 \ast \ww_2 = &(\ww_1 \odot \ww_2)'
    \\
    = &\frac{1}{2}\left(\ww_1 \otimes \ww_2 + \ww_2 \otimes \ww_1 \right) - \frac{1}{3} (\ww_1 \cdot \ww_2) \,\Idd,
  \end{split}
\end{equation*}
where $\ww_1 \cdot \ww_2=\ww_1^T \ww_2$ is the scalar product.

For two second-order harmonic tensors (deviators) $\bh_1$, $\bh_2$, we have
\begin{multline}\label{eq:H2astH2}
  \bh_1 \ast \bh_2 = \bh_1 \odot \bh_2 - \frac{2}{7}\, \Idd \odot (\bh_1 \bh_2 + \bh_2 \bh_1)
  \\
  + \frac{2}{35} \tr(\bh_1 \bh_2)\, \Idd \odot \Idd.
\end{multline}

Sylvester's theorem~\citep{Syl1909,Bac1970,Bae1998} states that any harmonic tensor $\bH$ of order $n$ can be factorized as
\begin{equation*}
  \bH= \vec w_1 \ast \vec w_2 \ast\dotsb \ast\vec w_n,
\end{equation*}
\textit{i.e.} as the harmonic products of $n$ (real) vectors $\vec w_k$, the so-called \emph{Sylvester-Maxwell multipoles}. Note however, that this factorization is far from being unique, as discussed in~\citep{OKDD2016}.

Setting $\bh_1= \vec w_1 \ast \vec w_2$ and $\bh_2= \vec w_3 \ast \vec w_4$ which are harmonic second-order tensors (deviators), we obtain the non unique harmonic factorization of $\bH$ by means of two second-order tensors:
\begin{equation}\label{eq:SylvesterTensorT2}
  \bH= \bh_1 \ast \bh_2,
\end{equation}
as detailed in~\citep{DD2016,OKDD2016}.

\section{Link between fourth-order crack density and damage tensors}
\label{sec:damage-tensors}

Before formulating our main result, Theorem~\ref{thm:Omegaavech2}, we summarize, in this section, the present \emph{state--of--the--art} in Continuum Mechanics leading to the representation of damage of cracked media by a fourth-order tensor~\citep{Cha1979}. We make an explicit link with the harmonic decomposition and we present, by comparison to the 2D case, the problem of representation of damage by second-order tensors in 3D.

\subsection{Crack density function and tensors}

The damage state of a microcracked material is classically defined by spatial arrangement, orientation and geometry of the cracks present at the microscale~\citep{Kac1972,LO1980,Lad1983,Ona1984,LC1985,Mur1988,Kac1993}. The crack density, related to any possible 3D direction defined by a unit vector $\nn$, refers to a dimensionless scalar property defined in a continuous manner at the Representative Volume Element scale as a spatial crack density function $\Omega=\Omega(\nn)$. Owing to the property $\Omega(\nn)= \Omega(-\nn)$, it is expressed by means of a totally symmetric tensor $\bF$ (the so-called \emph{fabric tensor}) of even order $n=2r$~\citep{Kan1984} as:
\begin{equation}\label{eq:DefF}
  \Omega(\nn) = \bF \scal (\nn \otimes \nn \otimes \dotsb \otimes \nn)
\end{equation}
where $\scal$ means the contraction over the $n$ subscripts. Note that $\Omega(\nn)$ corresponds to a homogeneous polynomial (see section~\ref{subsec:totally-symmetric-tensors})
\begin{equation*}
  \rh(n_1, n_2, n_3) = \bF (\nn, \nn, \dotsc , \nn).
\end{equation*}
The fabric tensor $\bF$, which is totally symmetric, can be determined as the least square error approximation of an experimental (measured) density distribution $\Omega(\nn)$, $\bF$ being thus solution of
\begin{equation*}
  \min_{\bF} \norm{\frac{4\pi}{2n+1} \bF\scal\Idd^{\odot n} - \int_{\norm{\xx}=1} \Omega(\nn)\, \nn^{\otimes n}\,{\rm d}S}^2,
\end{equation*}
with solid angle $S$ and where
\begin{equation*}
  \nn^{\otimes k} := \nn \otimes \nn \otimes \dotsb \otimes \nn.
\end{equation*}
Moreover, the following equality has been used:
\begin{equation*}
  \frac{1}{4 \pi} \int_{\norm{\xx}=1} \nn^{\otimes 2n}\,{\rm d}S = \frac{1}{2n+1} \Idd^{\odot n}.
\end{equation*}
Note that $\nn^{\otimes k}=\nn^{\odot k}$ is a totally symmetric tensor.

Comparative studies of the tensorial order, needed to represent given microcracking patterns, can be found in~\citep{LK1993,Kra1996,TNS2001}.

Expression~\eqref{eq:DefF} is often rewritten into the finite expansion~\citep{Kan1984,Ona1984,Kra1996}:
\begin{multline}\label{eq:Kanatani}
  \Omega(\nn) = \tq F_4 \scal (\nn \otimes \nn \otimes \nn \otimes \nn) + \bOmega_{6} \scal (\nn^{\otimes 6}) + \dotsb                                                                                     \\
  \dotsb + \bOmega_{2k} \scal (\nn^{\otimes 2k})+ \dotsb + \bOmega_{n} \scal (\nn^{\otimes n})
\end{multline}
with fourth-order part
\begin{equation}\label{eq:F4}
  \begin{aligned}
  \tq F_4\scal (\nn \otimes \nn \otimes \nn \otimes \nn) = &\Omega_{0}  + \bOmega_2 \scal (\nn \otimes \nn)
  \\ &+ \bOmega_4 \scal (\nn \otimes \nn \otimes \nn \otimes \nn)
    \end{aligned}
\end{equation}
with $n=2r$ even and where $\bOmega_{2k}$ are totally symmetric traceless (harmonic) tensors of order $2k$. The scalar term $\Omega_{0}$ is the crack density within considered Continuum Mechanics representative volume element
\begin{equation*}
  \Omega_{0} = \frac{1}{4\pi}\int_{\norm{\xx}=1} \Omega(\nn)\, {\rm d}S.
\end{equation*}
Crack density tensors $\bOmega_{2}, \bOmega_4, \dotsc, \bOmega_n$ are harmonic tensors of even order $2, 4, \dotsc, n$. They constitute independent crack density variables representative of the microcraking pattern (and anisotropy), determined uniquely up to order $n$ from the knowledge of the 3D spatial crack density distribution $\Omega(\nn)$.

\subsection{Derivation of the crack density tensors from the harmonic decomposition}
\label{sec:crack-density-tensors}

Let us point out that the harmonic tensors $\bOmega_{2k}$ correspond to the tensors $\bH_{r-k}$ issued from the harmonic decomposition~\eqref{eq:tensorial-symmetric-harmonic-decomposition} of the fabric tensor $\bF$:
\begin{equation*}
  \bF = \bH_{0} + \Idd \odot \bH_{1} + \dotsb + \Idd^{\odot r-1}\odot \bH_{r-1}+ \Idd^{\odot r}\, H_{r},
\end{equation*}
with $r=n/2$, where $H_{r} = H_{\frac{n}{2}}$ and the harmonic tensors $\bH_{k}$ of degree $n-2k$ are given by~\eqref{eq:Hr} and~\eqref{eq:Hk}. Observe, moreover, that:
\begin{equation*}
  (\Idd^{\odot k}\odot \bH_{k}) \scal \nn^{\otimes n} = \bH_{k} \scal \nn^{\otimes n-2k},
\end{equation*}
and we get thus:
\begin{multline*}
  \Omega(\nn)= H_r + \bH_{r-1} \scal ( \nn\otimes \nn)
  \\
  + \bH_{r-2} \scal ( \nn\otimes \nn \otimes \nn\otimes \nn) + \dotsb + \bH_{0} \scal (\nn^{\otimes n}),
\end{multline*}
which is the finite expansion~\eqref{eq:Kanatani}, where
\begin{equation*}
  \Omega_{0}=H_{\frac{n}{2}}, \qquad \bOmega_{2k}=\bH_{r-k}.
\end{equation*}

\subsection{Fourth-order damage tensor}
\label{S:FourtOrderDamage}

Using the decomposition~\eqref{eq:Kanatani} and assuming open microcracks in an initially 3D isotropic medium,~\cite{LO1980} and~\cite{Ona1984} have shown that the damage variable defined by the coupling microcraking/elasticity is at most a fourth-order tensor,
built from $\bF_4$ only, see~\eqref{eq:Kanatani}. This result holds for non interacting closed---sliding without friction---pennyshaped microcracks~\citep{Kac1993} and, as pointed out by~\cite{CW2010}, for many stress based homogenization schemes, as long as all the microcracks are in the same state, either open or closed. Setting:
\begin{equation*}
  \tq J = \mathbf{I}-\frac{1}{3} \Idd \otimes \Idd,
\end{equation*}
the following general definition of a fourth-order damage tensor has then been derived for initially isotropic materials ~\citep{Kac1993,ZC1998,CW2010}:
\begin{multline}\label{eq:D}
  \bD = p_{0} \Omega_{0} \Idd \otimes \Idd + p_1\Omega_{0} \, \tq J + p_2 (\Idd \otimes \bOmega_2+ \bOmega_2 \otimes \Idd)
  \\
  + p_3( \Idd \otimesbar \bOmega_2 + \bOmega_2 \otimesbar \Idd)+ p_4 \bOmega_4,
\end{multline}
where $\Omega_{0}$ should be interpreted as a scalar damage variable, the symmetric deviator $\bOmega_2$ as a second-order damage variable, and the harmonic tensor $\bOmega_4$ as a fourth-order damage variable. The expression of the scalars $p_{i}$ depends on the initial elasticity parameters, on the homogenization scheme and on the microcracks state (simultaneously open or simultaneously closed for all cracks).

\begin{rem}\label{rem:indep-Omega}
  The scalars $p_{i}$ do not depend on $\Omega_{0}, \bOmega_2, \bOmega_4$.
\end{rem}

\begin{rem}\label{rem:propD}
  \eqref {eq:D} is the harmonic decomposition~\eqref{eq:HarmE} of the fourth-order damage tensor $\bD$, which has the major and the minor indicial symmetries ($D_{ijkl}=D_{klij}=D_{jikl}$) as an elasticity tensor. The deviatoric parts of the dilatation and Voigt tensors are both proportional to the second-order harmonic tensor $\bOmega_2$, with the scalar factors $\kappa_{\textrm{di}}$ and $\kappa_{\textrm{vo}}$ depending only on the initial elastic parameters of the undamaged isotropic material:
  \begin{align*}
    \td {di}'(\bD) & = (\tr_{12} \bD)' = \kappa_{\textrm{di}} \bOmega_2, \\
    \td {vo}'(\bD) & = (\tr_{13} \bD)' = \kappa_{\textrm{vo}} \bOmega_2.
  \end{align*}
  The traces of the dilatation and the Voigt tensors are both proportional to the scalar crack density $\Omega_{0}$, with scalar factors
  $k_{\textrm{di}}$ and $k_{\textrm{vo}}$ depending only on the elastic parameters of virgin (undamaged) isotropic material:
  \begin{align*}
    \tr \td {di}(\bD) & =  \tr (\tr_{12} \bD) =  k_{\textrm{di}} \Omega_{0}, \\
    \tr \td {vo}(\bD) & =  \tr (\tr_{13} \bD) =  k_{\textrm{vo}} \Omega_{0}.
  \end{align*}
\end{rem}

{
\begin{rem}
An alternative framework is due to \cite{VK2006}. 
These authors extend to anisotropic damage the framework of \cite{ZC1995} for the representation of microstructure morphology of granular materials (the considered framework neglects the fourth order contribution $\bOmega_4$).
They propose, then, a nonlinear link between the  second order fabric tensor $\bOmega_2$ and a fourth order tensorial damage variable $\tq D$, setting for the effective (damaged) elasticity tensor 
\beq\label{eq:VKfromZC}
\begin{split}
	\tilde {\tq E} = \tilde {\tq E} =\lambda\,  \bphi \otimes \bphi + 2\mu \, \bphi\otimesbar \bphi, 
	\quad\,
	\bphi=(\Omega_0 \, \Idd + \bOmega_2)^{-k}
\end{split}
\eeq
This means that they replace the identity tensor $\Idd$ in the usual isotropic elasticity law by a negative power $-k=-0.2$ 
of the second order crack density tensor $\Omega_0 \Idd + \bOmega_2$
(powers being taken in terms of the principal values, $\lambda$, $\mu$ considered as Lam\'e-like constants). The crack density $\Omega_0$ must have a non zero initial value. Due to the presence of quadratic terms in crack densities, this framework does not satisfy previous proportionality properties.
\\ 
Fourth order damage tensor $\tq D$ is such that $\tilde {\tq E} =(\Idd -\tq D): \tq E$. 
\end{rem}
}

\subsection{2D case}
\label{subsec:2D-case}

In 2D, cracks are represented by 2D straight lines. Expression~\eqref{eq:Kanatani} for crack density holds, recovering a Fourier finite expansion~\citep{Kan1984,BHL1995}:
\begin{multline}\label{eq:Omega2D}
  \Omega(\nn)= \omega_{2D}+ \bomega'_{2D}\scal (\nn \otimes \nn)
  \\
  + \bH_{2D} \scal (\nn \otimes \nn \otimes \nn \otimes \nn) + \dotsb,
\end{multline}
where the unit vector $\nn$ is related to the possible planar direction
\begin{equation*}
  \omega_{2D}=\frac{1}{2\pi}\int_{0}^{2\pi} \Omega(\nn)\, {\rm d} \theta
\end{equation*}
is the 2D crack density, and where $\bomega'_{2D}$, $\bH_{2D}$ are respectively the 2D harmonic second and the fourth-order crack density tensors.

Verchery's decomposition~\citep{Ver1979,Van2005}, and its rewriting into a tensorial form~\citep{DD2015}, shows that any 2D harmonic fourth-order tensor is an harmonic square. Applied to $\bH_{2D}$, this gives:
\begin{equation}\label{eq:H2Dsquare}
  \bH_{2D} = \bh_{2D}\ast \bh_{2D},
\end{equation}
where $\bh_{2D}$ is an harmonic second-order tensor (deviator), and the notation $\ast$ denotes the harmonic product defined as the orthogonal projection of the symmetrized product $\bh_{2D} \odot \bh_{2D}$ onto 2D fourth-order harmonic tensors' space:
\begin{equation*}
  \bh_{2D}\ast \bh_{2D} := (\bh_{2D} \odot \bh_{2D})_{0}.
\end{equation*}
For second-order harmonic tensors, this reads:
\begin{equation*}
  \bh_{2D}\ast \bh_{2D}=\bh_{2D}\odot \bh_{2D}-\frac{1}{4}(\tr \bh_{2D}^2)\,\Idd \odot \Idd.
\end{equation*}

This means that in 2D, any anisotropic microcracking pattern can be expressed, up to order $4$, exactly by means of the scalar $\omega_{2D}$ and the two independent second-order deviatoric damage variables $\bomega'=\td \Omega_2^{2D}$ and $\bh_{2D}=\bh_{2D}'$. This result is consistent with the fact that the micro-mechanics of 2D media with open and closed (sliding without friction) microcracks can be represented by two second-order damage tensors only~\citep{DD2015}.

The question arises then whether the expansion \eqref{eq:Omega2D}--\eqref{eq:H2Dsquare} holds in 3D, \textit{i.e.} with $\bomega'$ and $\bh$ (now 3D) deviatoric second-order tensors. The answer is negative in the general triclinic case for which we have only $\bOmega_4=\bh_1 \ast \bh_2$ \eqref{eq:SylvesterTensorT2} with usually different second-order tensors $\bh_1, \bh_2$~\citep{OKDD2016}. Furthermore, the factorization is not unique, forbidding to interpret $\bh_1$ and $\bh_2$ as damage variables.

\section{{3D second-order damage tensors from walled structures}}
\label{sec:3D-crack-density}

It was noticed by~\cite{LK1993} and~\cite{Kra1996} that the fourth-order crack density tensor $\bOmega_4=(\bF_4)_{0}$ (the harmonic part of the totally symmetric tensor $\bF_4$) induced by particular loadings is related, sometimes, as a square of the second-order harmonic contribution $\bOmega_2$. More precisely, $\bOmega_4$ is proportional to the harmonic square $\bOmega_2 \ast \bOmega_2$ in the particular situation that occurs for a family of parallel penny shaped microcracks having identical---therefore coplanar---normal $\vec m$ (induced for example by uniaxial tension on quasi-brittle materials). 

{An harmonic square is also present in the work of \cite{VK2006}. Let us show that it corresponds to the fourth order harmonic tensor $\bphi \ast \bphi$ with $\bphi=(\Omega_0 \, \Idd + \bOmega_2)^{-0.2}$:
\begin{itemize}
\item[--]  The harmonic part $(\tilde {\tq E})_0$ of the effective elasticity tensor \eqref{eq:VKfromZC} is
\beq
	(\tilde {\tq E})_0= \left( \lambda\, (\bphi \otimes \bphi)^\mathrm{s}+2\mu\, (\bphi \otimesbar \bphi)^\mathrm{s}\right)_0,
\eeq
where $(.)^\mathrm{s}$ means the totally symmetric part and $(.)_0$ is the harmonic projection defined in Eq. \eqref{eq:E0}.
\item[--]   By definition $(\bphi \otimes \bphi)^\mathrm{s}=(\bphi\otimesbar \bphi)^s = \bphi\odot \bphi$.
\item[--]   The harmonic fourth order part of $\tilde {\tq E}$ is thus 
\begin{equation*}
  (\tilde {\tq E})_{0} = ( \lambda+2\mu )(\bphi\odot \bphi)_{0} =( \lambda+2\mu)\, \bphi\ast \bphi,
\end{equation*}
where $\ast$ is the harmonic product \eqref{eq:H2astH2}. See also \ref{A:HarmDecompD}.
\end{itemize}
}

{The second and fourth-order crack density variables $\bOmega_2$ and $\bOmega_4$ are independent in general case. We prefer to keep the fourth order contribution $\bOmega_4$ in our further analyses and we will use next the standard expression \eqref{eq:D} (for instance instead of Eq. \eqref{eq:VKfromZC} which neglects $\tq \Omega_4$).
}

{
In order to built a micromechanics based framework with second-order damage tensors, one considers as the general case---except for soils---that the measurements of 3D crack density $\Omega(\nn)$ is performed on thin {or thick} walled structures, \textit{i.e.} on thin or thick tubes or on 2D structures such as plates. }This allows us to introduce the unit normal $\bnu$ ($\norm{\bnu}=1$) to the walled structure and the set
\begin{equation*}
  \MES(\bnu):=\set{\pmb{\tau},\norm{\pmb{\tau}}=1 \; \textrm{and} \; \pmb{\tau}\cdot\bnu=0} \cup \{\bnu\},
\end{equation*}
of directions $\nn$ restricted to so-called \emph{mechanically accessible directions for measurements}. 

As an extension of both the previous remark on fourth-order harmonic squares and the 2D result~\eqref{eq:Omega2D}--\eqref{eq:H2Dsquare},
we have in 3D the following theorem (the proof of which is given at the end of the present section):

\begin{thm}\label{thm:Omegaavech2}
  For a given unit vector $\pmb{\nu}$, any density function $\Omega(\nn)$ is represented, up to fourth-order, for all directions $\nn \in \MES(\bnu)$ by means of a scalar $\omega_m$ and two harmonic (symmetric deviatoric) second-order tensors $\bomega'$ and $\bh$ as:
  \begin{multline}\label{eq:Dden3D}
    \Omega(\nn) =\omega_m +\bomega' \scal (\nn \otimes \nn) \\
    + \left(\bh\ast \bh \right) \scal (\nn \otimes \nn \otimes \nn \otimes \nn) + \dotsb
  \end{multline}
  for all $\nn \in \MES(\bnu)$. This representation is unique, up to $\pm\bh$, if $\left.(\bomega' \bnu) \times \bnu= \bh\, \bnu=\vec 0\right.$.
\end{thm}

\begin{rem}
  If we set $\vec e_3=\bnu$, the conditions $(\bomega' \bnu) \times \bnu=\vec 0$ (which is equivalent to $(\bomega' \bnu) = \lambda\, \bnu$) and $\bh\, \bnu=\vec 0$ mean that
  \begin{equation}\label{eq:matrix-omega}
    \bomega'  =
    \begin{pmatrix}
      \omega_{11}' & \omega_{12}  & 0                              \\
      \omega_{12}  & \omega_{22}' & 0                              \\
      0            & 0            & -(\omega_{11}' +\omega_{22}' )
    \end{pmatrix},
  \end{equation}
  and
  \begin{equation}\label{eq:matrix-h}
    \bh  =
    \begin{pmatrix}
      h_{11} & h_{12}   & 0 \\
      h_{12} & - h_{11} & 0 \\
      0      & 0        & 0
    \end{pmatrix}.
  \end{equation}
\end{rem}

Applied to thinned walled structured, for which $\nn\in \MES(\bnu)$ are the accessible directions for mechanical measurements, Theorem \ref{thm:Omegaavech2} states that any microcracking pattern, possibly triclinic, can be represented, up to order $4$, by means of only two symmetric second-order crack density tensors, $\bomega=\bomega'+\omega_m\, \Idd$ and $\bh$, the second one being a deviator. In that case, we can recast~\eqref{eq:Dden3D} as:
\begin{equation*}
  \Omega(\nn) = \,\bomega \scal (\nn \otimes \nn) + \left(\bh\ast \bh \right) \scal (\nn \otimes \nn \otimes \nn \otimes \nn) + \dotsb
\end{equation*}
for all $\nn \in \MES(\bnu)$, where $\bomega$ is the crack density tensor introduced by~\cite{VK1971}~\citep{Kac1972,Kac1993}.

\subsection*{Practical formulas -- Proof of Theorem \ref{thm:Omegaavech2}}
\label{S:proofOmegaavech2}

Recall that, up to order four, the crack density function $\Omega(\nn)$ is represented by the fabric tensor $\bF_4$~\eqref{eq:Kanatani}--\eqref{eq:F4}. Consider now an orthonormal frame $\set{\vec e_1, \vec e_2, \bnu}$ and let $(\omega_m, \bomega', \bh)$ be a triplet as in~\eqref{eq:matrix-omega}--\eqref{eq:matrix-h}. Set
\begin{multline}\label{eq:hij}
  h_{11} +i h_{12} = \frac{1}{2} \big[(\bF_4)_{1111}+(\bF_4)_{2222} - 6(\bF_4)_{1122}
    \\
  + 4 i \big((\bF_4)_{1112}-(\bF_4)_{1222}\big)\big]^{1/2}
\end{multline}
and
\begin{equation*}
  \begin{aligned}
    \omega_m=     & \frac{1}{4}(\bF_4)_{1111}+\frac{1}{2} (\bF_4)_{1122}+\frac{1}{4} (\bF_4)_{2222}
    \\
                  & +\frac{1}{3} (\bF_4)_{3333}-\frac{1}{15}(h_{11}^2+h_{12}^2),
    \\
    \omega_{11}'= & \frac{5}{8} (\bF_4)_{1111}+\frac{1}{4} (\bF_4)_{1122}-\frac{3}{8} (\bF_4)_{2222}
    \\
                  & -\frac{1}{3} (\bF_4)_{3333}+\frac{1}{42} (h_{11}^2+ h_{12}^2),
    \\
    \omega_{22}'= & -\frac{3}{8} (\bF_4)_{1111}+\frac{1}{4} (\bF_4)_{1122}+\frac{5}{8} (\bF_4)_{2222} \\
                  & -\frac{1}{3} (\bF_4)_{3333}+\frac{1}{42}(h_{11}^2+h_{12}^2),
    \\
    \omega_{12}=  & (\bF_4)_{1222}+(\bF_4)_{1112},
  \end{aligned}
\end{equation*}
where $i=\sqrt{-1}$ is the pure imaginary number. It can be checked by a direct computation that $(\omega_m, \bomega', \bh)$ is a solution of \eqref{eq:Dden3D} in Theorem~\ref{thm:Omegaavech2}.

\begin{rem}\label{rem:pmh}
  Because of the square root in~\eqref{eq:hij}, both $\bh$ and $-\bh$ are solutions.
\end{rem}

We will now show the uniqueness of the solution, up to a sign, and under the assumption that:
\begin{equation*}
  (\bomega' \bnu) \times \bnu= \bh\, \bnu=\vec 0.
\end{equation*}
Suppose thus, that a second solution $(\omega_m^*, \bomega^{*\prime}, \bh^*)$ to~\eqref{eq:Dden3D} exists, with $\bomega^{*\prime}$ and $\bh^*$ as in~\eqref{eq:matrix-omega}--\eqref{eq:matrix-h}. Equaling, for different directions $\nn\in \mathcal R(\bnu)$, the density function $\Omega(\nn)$ defined by~\eqref{eq:Dden3D}, calculated first with $(\omega_m,\bomega^{\prime},\bh)$ and then with $(\omega_m^*,\bomega^{*\prime},\bh^*)$, we obtain for $\nn=\bnu$:
\begin{multline}\label{eq:3333}
  35(\omega^{*\prime}_{11} + \omega^{*\prime}_{22} - \omega'_{11} - \omega'_{22} + \omega_m - \omega_m^*)
  \\
  + 4 h_{11}^2 + 4 h_{12}^2 - 4 h_{11}^{*2} - 4 h_{12}^{*2} = 0.
\end{multline}
Then, for $\nn=(\cos \theta, \sin \theta, 0)$, we get:
\begin{equation}\label{eq:Fourieruniq}
  a_{0} + a_{2}\cos 2 \theta + b_{2}\sin 2 \theta + a_{4}\cos 4 \theta + b_{4}\sin 4 \theta = 0,
\end{equation}
where
\begin{align*}
  a_{0} & = 2 (\omega_m-\omega_m^*)+\omega'_{11}-\omega^{*\prime}_{11}+\omega'_{22}-\omega^{*\prime}_{22} \\
        & \quad + \frac{3}{35} (h_{11}^2-h_{11}^{*2}+h_{12}^2-h_{12}^{*2}),                               \\
  a_{2} & = \omega'_{11}-\omega^{*\prime}_{11}-\omega'_{22}+ \omega^{*\prime}_{22},                       \\
  b_{2} & = 2 (\omega_{12}-\omega^*_{12}),                                                                \\
  a_{4} & = h_{11}^2-h_{12}^2-h_{11}^{*2}+h_{12}^{*2},                                                    \\
  b_{4} & = 2 (h_{11} h_{12}-h^*_{11} h^*_{12}).                                                          \\
\end{align*}
Since~\eqref{eq:Fourieruniq} holds for all $\theta$, we have
\begin{equation*}
  a_{0} = a_{2} = b_{2} = a_{4} = b_{4} = 0.
\end{equation*}
Since $a_{4} = b_{4} = 0$, we get
\begin{equation*}
  \left\{
  \begin{aligned}
    h_{11}^{*2}-h_{12}^{*2}= & h_{11}^2-h_{12}^2,
    \\
    h^*_{11} h^*_{12} =      & h_{11} h_{12},
  \end{aligned}
  \right.
\end{equation*}
from which we deduce that
\begin{equation*}
  (h^*_{11}+i h^*_{12})^2= (h_{11}+i h_{12})^2
\end{equation*}
and therefore that $\bh^*=\pm \bh$ (in accordance with Remark \ref{rem:pmh}). From~\eqref{eq:3333} and $a_{0} = a_{2} = b_{2} = 0$, we get
\begin{equation*}
  \left\{
  \begin{aligned}
    \omega_m-\omega_m^*+\omega^{*\prime}_{11}-\omega'_{11}+\omega^{*\prime}_{22}-\omega'_{22}     & = 0,
    \\
    2 (\omega_m-\omega_m^*)+\omega'_{11}-\omega^{*\prime}_{11}+\omega'_{22}-\omega^{*\prime}_{22} & = 0,
    \\
    \omega'_{11}-\omega^{*\prime}_{11}-\omega'_{22}+
    \omega^{*\prime}_{22}                                                                         & = 0,
    \\
    \omega_{12}-\omega^*_{12}                                                                     & = 0,
  \end{aligned}
  \right.
\end{equation*}
\textit{i.e.} $\omega_m^*=\omega_m$ and $\bomega^{*\prime}=\bomega^{\prime}$, which achieves the proof.

\section{General micro-mechanics based framework with two second-order damage variables}
\label{sec:general-micro-based}

Using the results from Section~\ref{S:FourtOrderDamage}, we deduce from~\eqref{eq:Dden3D} that the representation by means of two symmetric second-order tensors holds for the damage tensor itself, at least when the microcracks are all in the same state, all open or all closed. This means that, disposing from sufficiently many in-plane measurements (along directions $\nn$ orthogonal to $\bnu$) and an out-of-plane measurement (along $\nn=\bnu$), the general fourth-order damage tensor of Chaboche--Leckie--Onat can be expressed by means of two symmetric second-order damage variables only, for example $\bomega$ and $\bh$ (the second--one being a deviator). A general damage framework using this feature is derived next, clarifying the link between~\cite{CS1982} and~\cite{Lad1983,Lad1995} phenomenological second-order damage models and micro-mechanics based framework.

We shall assume that the homogenization result~\eqref{eq:D} holds, where the constants $p_{i}$ are given (refer to the works of~\cite{Kac1993} and~\cite{DK2016} for comparison of different homogenization schemes). Gibbs free enthalpy density writes
\begin{equation*}
  \rho \psi^\star = \frac{1}{18K} (\tr \bsigma)^2+\frac{1}{4G} \bsigma':\bsigma' + \frac{1}{2E}\, \bsigma: \bD:\bsigma,
\end{equation*}
where $\rho$ is the density and $E$, $G=\frac{E}{2(1+\nu)}$, $K=\frac{E}{3(1-2\nu)}$ are, respectively, the Young, shear and bulk moduli.
The elasticity law, coupled with the anisotropic damage, writes then as
\begin{equation*}
  \pmb \epsilon^e = \rho \frac{\partial \psi^\star}{\partial \bsigma} = \frac{1}{2G} \bsigma' + \frac{1}{9K} (\tr
  \bsigma)\, \Idd + \frac{1}{E}\, \bD:\bsigma,
\end{equation*}
or, in a more compact form, as
\begin{equation*}
  \pmb \epsilon^e = \tilde{\tq S}:\bsigma ,
\end{equation*}
where $\pmb \epsilon^e$ is the elastic strain tensor and $\tilde {\tq S}$, the effective fourth-order compliance tensor
\begin{equation} \label{eq:EffecitiveComplianceAndDamageTensor}
  \tilde{\tq S} = \frac{1}{9K} \Idd \otimes \Idd +\frac{1}{2G} \tq J + \frac{1}{E}\, \bD,
  \quad
  \tq J = \mathbf{I}-\frac{1}{3} \Idd \otimes \Idd .
\end{equation}

Having many in-plane and possibly one out-of-plane measurements, allows us to use remark~\ref{rem:indep-Omega} and \eqref{eq:Dden3D} instead of \eqref{eq:D}, within the considered homogenization scheme. We can thus recast the fourth-order damage tensor $\bD$ by substituting the scalar $\Omega_{0}$ by $\omega_m$, the second-order tensor $\bOmega_2$ by the deviatoric tensor $\bomega'$ and the fourth-order tensor $\bOmega_4$ by the harmonic (\textit{i.e} totally symmetric and traceless) tensor $\bh \ast \bh$. More precisely, we get
\begin{multline}\label{Omegahh}
  \bD = p_{0} \omega_m\, \Idd \otimes \Idd + p_1 \omega_m\, \tq J + p_2\, (\Idd \otimes \bomega'+ \bomega' \otimes \Idd)
  \\
  + p_3\,( \Idd \otimesbar \bomega' + \bomega' \otimesbar \Idd) + p_4\,\bh\ast \bh .
\end{multline}
Using \eqref{eq:H2astH2}, the term $\bh \ast \bh$ expands as
\begin{equation*}
  \begin{split}
    \bh \ast \bh = & \frac{1}{3} \bh\otimes \bh + \frac{2}{3} \bh\otimesbar \bh
    \\
    &- \frac{2}{21}\,\left(
    \Idd \otimes \bh^2 + \bh^2 \otimes \Idd + 2 (\Idd \otimesbar \bh^2 + \bh^2 \otimesbar \Idd) \right)
    \\
    & + \frac{2}{105} (\tr\bh^2)\, (\Idd \otimes \Idd + 2\; \Idd \otimesbar \Idd),
  \end{split}
\end{equation*}
so that the enthalpic contribution, due to the microcracks, writes
\begin{equation*}
  \begin{aligned}
    \bsigma:\bD:\bsigma & =  p_{0} \omega_m \, (\tr \bsigma)^2 + p_1 \omega_m \, \tr (\bsigma^{\prime 2})
    \\
                        & \quad + 2 p_2 \tr (\bomega' \bsigma) \tr \bsigma + p_3\tr (\bomega' \bsigma^2)
    \\
                        & \quad + p_4 \Big[ \frac{1}{3}  (\tr (\bh\bsigma'))^2 + \frac{2}{3} \tr(\bsigma' \bh \bsigma' \bh )
    \\
                        & \quad - \frac{8}{21}\tr ( \bh^2 \bsigma^{\prime 2})
    + \frac{4}{105} \tr\bh^2\, \tr \bsigma^{\prime 2}\Big].
  \end{aligned}
\end{equation*}

Again, as in Remark~\ref{rem:propD}, \eqref{Omegahh} is nothing else but the harmonic decomposition of the fourth-order damage tensor $\bD$, but with the following particularities. Let
\begin{equation*}
  \td{di}(\bD) := (\tr_{12} \bD)
\end{equation*}
be the \emph{dilatation tensor} of $\bD$ and
\begin{equation*}
  \td {vo}(\bD) := (\tr_{13} \bD)
\end{equation*}
be the \emph{Voigt tensor} of $\bD$. Then:
\begin{enumerate}
  \item[--]  the harmonic part $\bH=\bh \ast \bh$ of $\bD$ is factorized as an harmonic square;

  \item[--]  the following proportionality relations hold between the \emph{deviatoric parts} of $\td{di}(\bD)$ and $\td {vo}(\bD)$:
        \begin{equation}\label{trijDprim}
          \td{di}'(\bD) \, \propto \, \td {vo}'(\bD) \, \propto\, \bomega',
        \end{equation}
        which is equivalent for the effective compliance tensor $\tilde{\tq S}$ to satisfy the same conditions:
        \begin{equation}\label{trijtildeSprim}
          \td{di}'(\tilde {\tq S}) \, \propto \, \td {vo}'(\tilde {\tq S}) \, \propto\, \bomega' ;
        \end{equation}

  \item[--]  the following proportionality relations hold between the \emph{traces} of $\td {di}(\bD)$ and $\td {vo}(\bD)$:
        \begin{equation}
          \tr \td {di}(\bD) \, \propto \, \tr \td {vo}(\bD) \, \propto \, \omega_m .
        \end{equation}
\end{enumerate}

Following~\cite{CW2010}, who consider the scalar constants $p_{i}$ as material parameters, \emph{conditions 1 to 3 above, are the conditions for a damage model}---for instance built in a phenomenological manner---\emph{which should be considered as micro-mechanics based.}

\section{A second-order anisotropic damage model in micro-mechanics based framework}
\label{S:2ndOFrame}

Following~\cite{CS1982} and~\cite{Lad1983}, a symmetric second-order, unbounded damage variable $\bPhi$ is introduced in the Gibbs free enthalpy (with initial value $\bPhi=\Idd$ for a virgin material, and with damage growth $\frac{\rm d}{{\rm d} t} \bPhi$ positive definite). The usual second-order damage tensor writes as:
\begin{equation*}
  \bd = \Idd - \bPhi^{-2} \quad (\text{with initial value $\bd=\pmb 0$}).
\end{equation*}
A general but phenomenological coupling of elasticity with second-order anisotropic damage is described in~\citep{Des2006}. It
reads
\begin{equation}\label{eq:Gibbsg}
  \rho \psi^\star= \frac{g(\bPhi)}{18K} (\tr \bsigma)^2+\frac{1}{4G} \tr(\bPhi\, \bsigma'\,\bPhi\, \bsigma' ),
\end{equation}
where $\bsigma' = \bsigma -\frac{1}{3} (\tr \bsigma) \Idd$ is the stress deviator. The function $g$ was chosen as
\begin{equation}\label{eq:3DH}
  g(\bPhi) := \frac{1}{1-\tr \bd}=\frac{1}{\tr \bPhi^{-2}-2}
\end{equation}
for metals in~\citep{LDS2000}, and as
\begin{equation*}
  g(\bPhi) := \frac{1}{3} \tr \bPhi^2
\end{equation*}
for concrete in~\citep{Des2016}. In both models, the convexity with respect to $\bsigma$ and the positivity of the intrinsic dissipation are satisfied~(see also \cite{CDR2014}).

{
The elasticity law writes
\begin{align} \label{eq:ElasticityLaw}
\bepsilon^e &= \rho \frac{\partial \psi^\star}{\partial \bsigma}
						= \frac{g(\bPhi)}{9K}(\tr \bsigma) \Idd + \frac{1}{2G} (\bPhi\, \bsigma'\,\bPhi)'
\end{align}
with effective compliance tensor
\begin{equation}\label{eq:ElastgCompl}
  \tilde {\tq S}= \frac{1}{9K} \Idd \otimes \Idd +\frac{1}{2G} \tq J + \frac{1}{E}\, \bD,
\end{equation}
where  $\frac{1}{E}\tq D$ has for harmonic decomposition (proof given in~\ref{A:HarmDecompD})
\begin{multline} \label{eq:ElastgComplD}
 \frac{1}{E}\bD= \frac{g(\bPhi)-1}{9K} \Idd \otimes \Idd
 +\frac{2 \beta -1}{2 G} \tq J
 \\
 -\frac{2}{3 G}(\Idd \otimes \bb'+ \bb' \otimes \Idd)
 \\
 + \frac{1}{G}(\Idd \otimesbar \bb' +\bb'  \otimesbar \Idd)
+ \frac{1}{2 G} \bPhi \ast \bPhi
\end{multline}
with
\begin{align*}
\beta&=\frac{1}{60}\left[ \tr (\bPhi^{2}) + 3 (\tr \bPhi)^{2}\right],
&
\td{b}' &=\frac{1}{14}\left[ 3 (\tr \bPhi) \bPhi'-(\bPhi^{2})'\right].
\end{align*}
}

{
The harmonic part of the fourth order damage tensor is the harmonic square,
\begin{equation}
(\tq D)_{0}= (1+\nu) \; \bPhi \ast \bPhi,
\end{equation}
where $\nu$ is Poisson ratio of undamaged material.
It satisfies the first condition on the effective compliance $\tilde {\tq S}$ to be of micro-mechanics based form \eqref{Omegahh}, with
\begin{equation*}
  p_4=1+\nu, \qquad \bh =\bPhi'.
\end{equation*}
}

{
Moreover, we have
\begin{align*}
\left(\tr_{12} \bD\right)' &=\pmb{0},
&
\left(\tr_{13} \bD\right)' &=\frac{7(1+\nu)}{3 } \bb'.
\end{align*}
Both deviatoric parts $\left(\tr_{12} \bD\right)'$ and $\left(\tr_{13} \bD\right)' $ are obviously proportional, so that the phenomenological anisotropic damage model 
satisfies the second condition on the effective compliance $\tilde {\tq S}$ to be of micro-mechanics based form \eqref{Omegahh}, with
\begin{align*}
     p_2 \bomega'=&
    \frac{2(1+\nu)}{21}\left((\bPhi^2)'-3 (\tr \bPhi)\; \bPhi' \right),
    \\
    p_3 \bomega'=& -\frac{3}{2}p_2 \bomega'.
\end{align*}
By identification of the isotropic part of the harmonic decomposition of $\tq D$, we get
\begin{align*}
    p_{0}\omega_m=&\frac{1-2\nu}{3}(g(\bPhi)-1),
    \\
    p_1\omega_m=& (1+\nu)\left(\frac{1}{10}(\tr \bPhi)^2+\frac{1}{30}\tr \bPhi^2 -1\right).
\end{align*}
Recall that the material constants $p_{0}, p_1$ are independent of $\bPhi$, they are considered as material parameters, so that the proportionality requirement $p_{0}\omega_m \,\propto \, p_1 \omega_m$ can be satisfied if---following~\cite{BHL1995} and~\cite{LDS2000}---we define the hydrostatic sensitivity parameter $\eta$ as the material constant}
\begin{equation*}
  \eta = \frac{3p_{0} (1+\nu)}{p_1(1-2\nu)},
\end{equation*}
and set
\begin{equation}\label{eq:geta}
  g(\bPhi)=(1-\eta)+\eta\left(\frac{1}{10}(\tr \bPhi)^2+\frac{1}{30}\tr \bPhi^2\right).
\end{equation}
Condition 3 for the model \eqref{eq:Gibbsg} to be considered as micro-mechanics based is then fulfilled. For $\eta\geq 0$, the Gibbs free enthalpy density (\ref{eq:Gibbsg}) is furthermore convex with respect to both the stress tensor $\bsigma$ and the damage tensor $\bPhi$.

{Note that a full anisotropic damage model---including damage evolution---for quasi brittle materials can be naturally derived following \citep{Des2016}. 
The hydrostatic sensitivity just obtained in Eq.~\eqref{eq:geta} is quantified in next Section.}

\section{{Hydrostatic sensitivity}}
\label{S:ShearBulk}

{
Let us consider quasi-brittle materials such as concrete and the micro-mechanically based damage model just derived (Section \ref{S:2ndOFrame}, Gibbs free enthalpy density (\ref{eq:Gibbsg}), elasticity law~\eqref{eq:ElasticityLaw} and function $g(\bPhi)$, given by Eq. \eqref{eq:geta}). The second order damage variable, which vanishes for virgin material, is
 $\bd = \Idd - \bPhi^{-2}$.
 }
 
{Even when the damage state is anisotropic, the constitutive  equations  considered in Section \ref{S:2ndOFrame} is such that the dilatation tensor of the effective compliance tensor is null.
This allows for the definition of an effective bulk modulus $\tilde K$ for damaged materials ({\it i.e.} a  modulus function of damage tensor and of material parameters of the undamaged material) such as
\begin{equation*}
	\tr \bsigma= 3 \tilde K \tr \bepsilon^e 
\end{equation*}
Calculating the trace of elasticity law~\eqref{eq:ElasticityLaw}, we get
\beq
	\tilde K= \frac{K}{ g(\bPhi)}
	= \frac{K}{(1-\eta)+\eta\left(\frac{1}{10}(\tr \bPhi)^2+\frac{1}{30}\tr \bPhi^2\right)},
	\label{eq:KtildeetaDH}
\eeq 
where $K=\frac{E}{3(1-2\nu)}$ is the bulk modulus for the virgin (undamaged) material. Recall that $g(\Idd)=1$.
}

\begin{figure}[htbp]
\begin{center}
\includegraphics[width= 40mm]{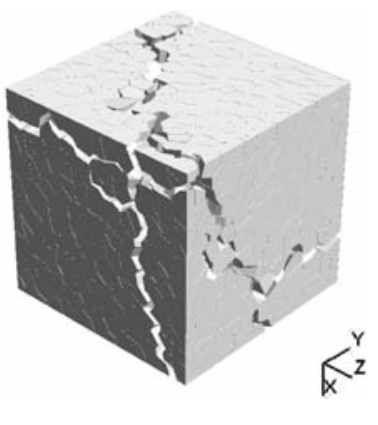}
\caption{Discrete Element sample  considered as Representative Volume Element ($16\times 16\times 16 \,\textrm{mm}^3$ cube, from \cite{DD2008}) and crack pattern at the end of equi-triaxial loading (high damage level).\label{fig:DEsamples} } 
\end{center}
\end{figure}

{
The measurement and therefore the identification of the material parameter $\eta$ are not easy tasks for quasi-brittle materials. This is why Discrete Elements \citep{CS1979,HR1990,SG1997,VM2002,OLDR2015} are often used as numerical experimentation for the tensile states of stresses of quasi-brittle materials. In such numerical tests the material is described as a particles assembly representative of the material heterogeneity, the particles being here linked by elastic-brittle beams. The size $16\times 16\times 16 \,\textrm{mm}^3$ of Representative Volume Element of a micro-concrete is considered (Fig. \ref{fig:DEsamples}); it is representative of a micro-concrete of Young's modulus $E=30000$ MPa and Poisson's ration $\nu=0.2$. The number of particles is 3,072
and the number of degrees of freedom 24,576.
The crack pattern obtained at the end of the equi-triaxial loading (quasi-rupture) is the one given in Fig. \ref{fig:DEsamples}, with a number of beams to break before failure of 8,000. 
}

\begin{figure}[htbp]
\begin{center}
\includegraphics[width=70mm]{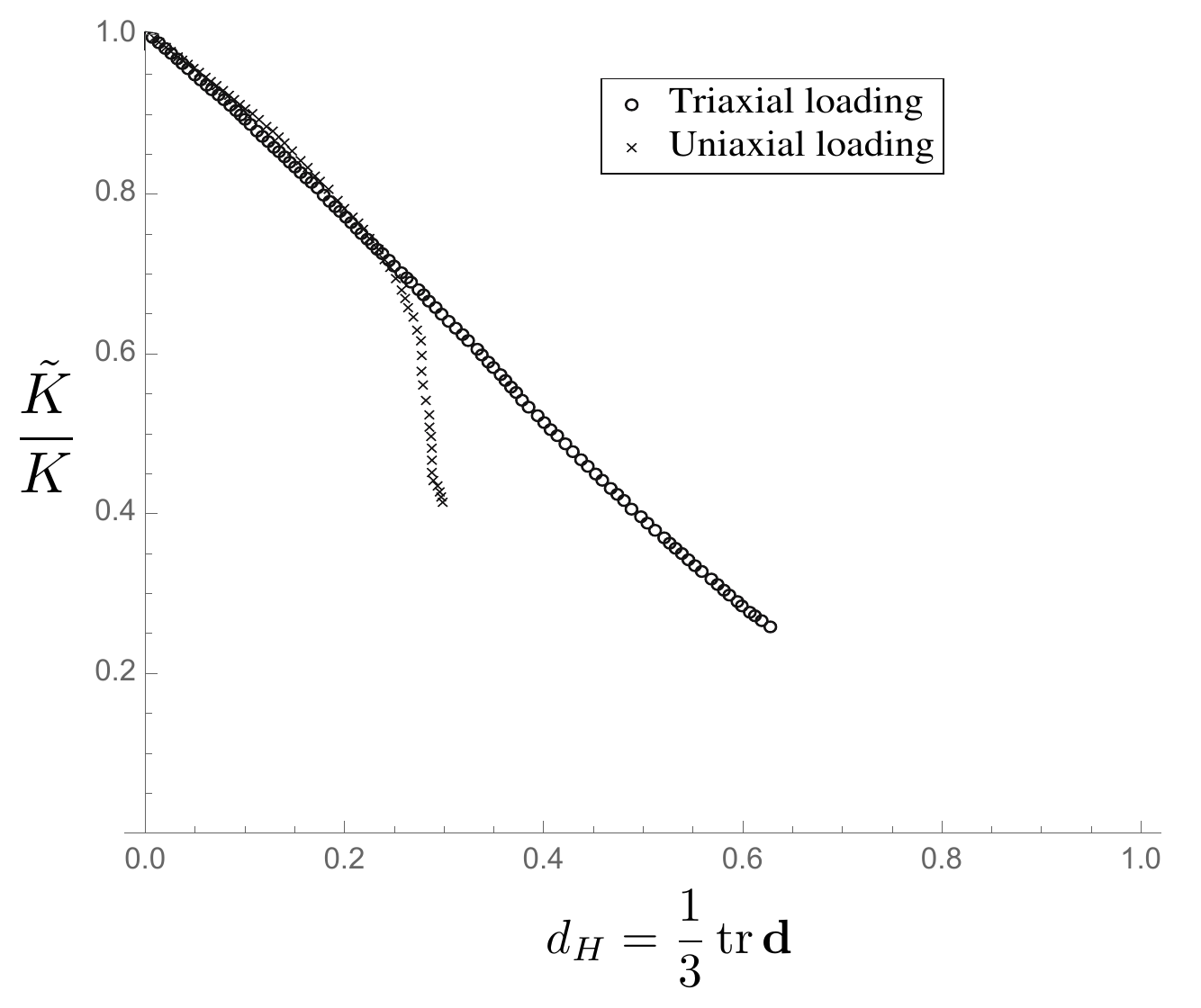}
\caption{Effective bulk modulus $\tilde K$ from Discrete Element computations as a function of hydrostatic damage $d_H$ (from \cite{DD2008}).
\label{fig:KtildeArnaud_DH}}   
\end{center}
\end{figure}

{
It has been shown from such computations (Fig. \ref{fig:KtildeArnaud_DH}) that at low damage the dependence $\tilde K(\textrm{damage})$ is close to be the same linear function 
\beq\label{eq:KtildedHDD}
	\tilde K \approx K (1- 1.2 \, d_H),
	\qquad
	d_H=\frac{1}{3} \tr \bd,
\eeq
of the hydrostatic damage $d_H$ whatever the stress triaxiality. For each mark of the Figures the components of the damage tensor $\bd= \Idd - \bPhi^{-2}$ have been measured by means of repeated numerical elastic loading-unloading sequences performed in uniaxial tension on the $16\times 16\times 16 \,\textrm{mm}^3$ cube (even for the triaxial loading), using then the coupling of elasticity with anisotropic damage given by elasticity law~\eqref{eq:ElasticityLaw} with one non zero principal stress $\sigma_i=\sigma$, the two others $\sigma_{j\neq i}=0$.
}

\begin{figure}[htbp]
\begin{center}
\includegraphics[width=70mm]{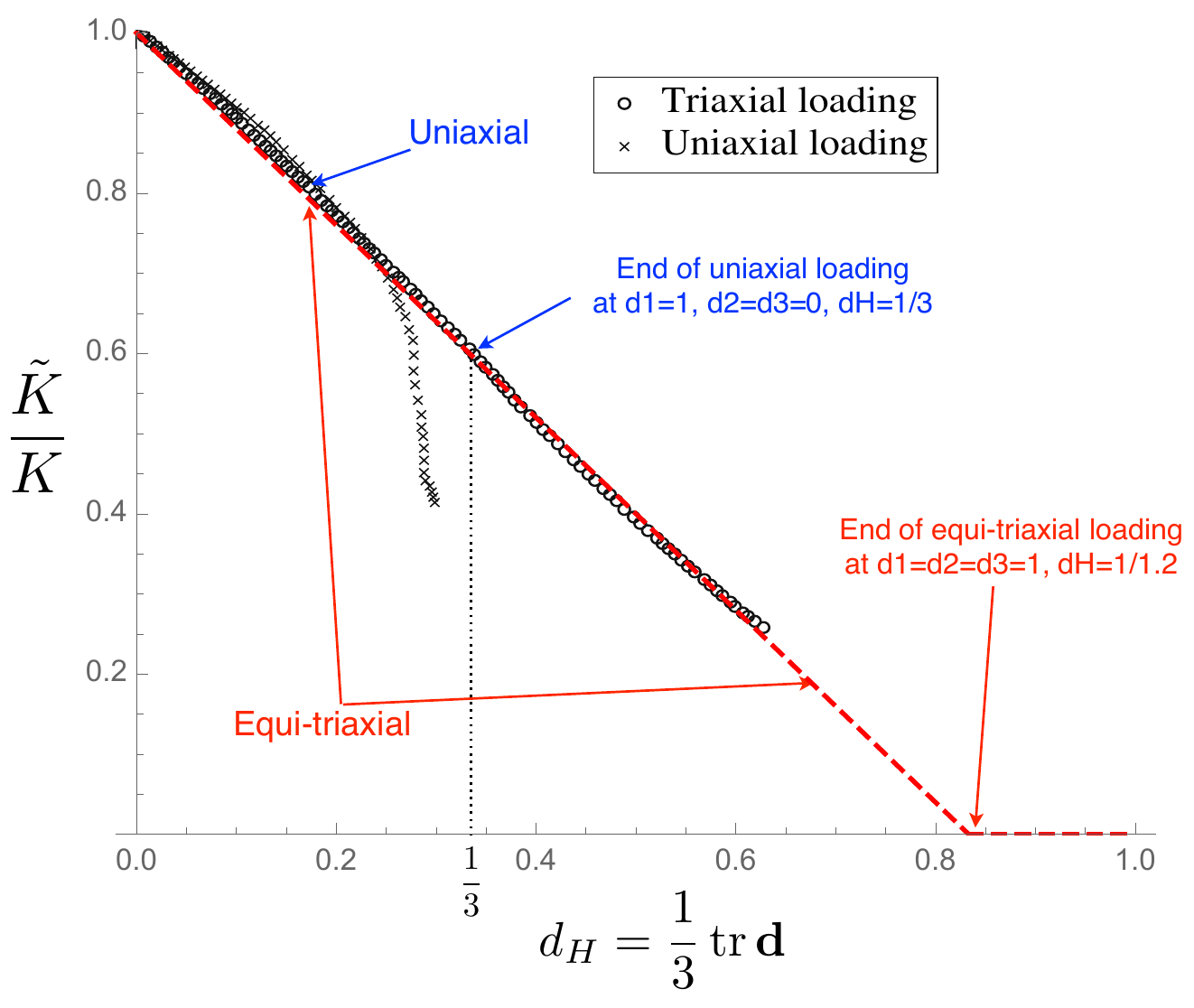}
\caption{Effective bulk modulus $\tilde K$ from initial modeling $\left.\tilde K=K(1-1.2\, d_H)\right.$ (dashed lines) 
as a function of hydrostatic damage.
\label{fig:KtildeArnaud_etadH}}   
\end{center}
\end{figure}

\begin{figure}[htbp]
\begin{center}
\includegraphics[width=70mm]{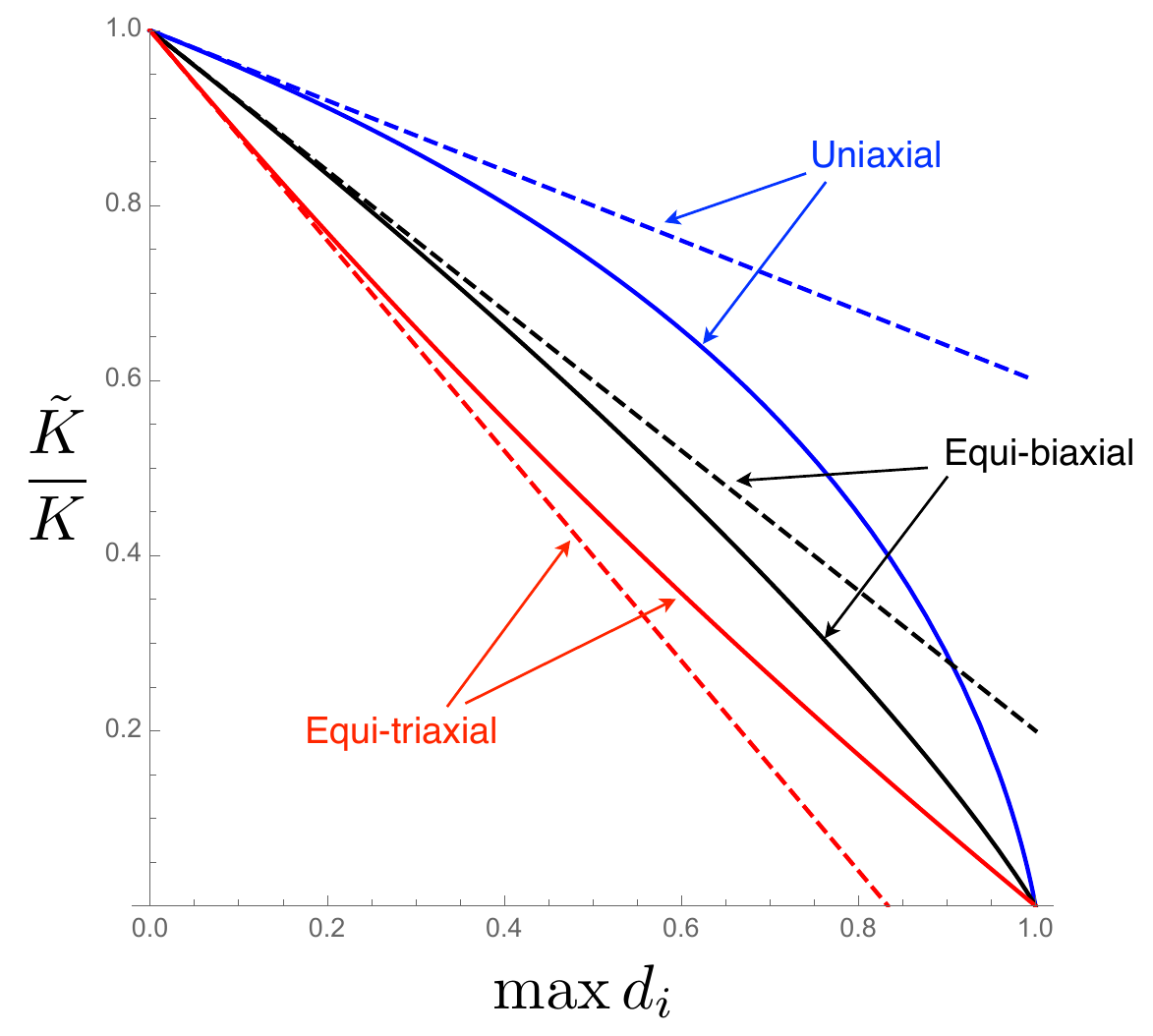}
\caption{Effective bulk modulus $\tilde K$ from~\eqref{eq:KtildeetaDH} (solid lines) and from initial modeling $\left.\tilde K=K(1-\eta\, d_H)\right.$ (dashed lines) as a function of hydrostatic damage $d_H$  with $\bd=1-\bPhi^{-2}$
(for value $\eta=1.2$ representative of a micro-concrete).
\label{fig:Ktildemaxdi}  }
\end{center}
\end{figure}

{
Note that considering Eq. \eqref{eq:KtildedHDD} at high damage level means that \emph{in uniaxial tension} the bulk modulus $\tilde K$ fully vanishes at $\tr \bd=3/1.2=2.5$, {\it i.e.} at maximum principal damage 
$\max d_i$ larger than 1  ($\tilde K$ cannot vanish then as principal damages $d_i$ ---therefore here $d_1$--- are always bounded by 1, see also Fig. \eqref{fig:Ktildemaxdi} for $\eta=1.2$). This corresponds to a quite high (spurious) elastic  stiffness $\tilde K$ which is kept at rupture (see Fig. \ref{fig:KtildeArnaud_etadH}).
Enforcing then gradually $\tilde K\rightarrow 0$ but allowing for damage tensor $\bd$ to evolve up to second order unit tensor $\Idd$ in an adequate  
procedure for the numerical control of rupture
is a solution which leads to numerical difficulties in Finite Element computations  \citep{BGL2007,RDG2008,Leroux12}.
}

{
The relation \eqref{eq:KtildeetaDH} relates the effective bulk modulus to  damage tensor $\bPhi$  and hydrostatic sensitivity parmeter $\eta$ (obtained from proposed micro-mechanically based second order damage framework by function $g$ defined in Eq. \eqref{eq:geta}).
This relation 
implies that the effective (damaged) bulk modulus $\tilde K$ vanishes exactly when the maximum eigenvalue of damage tensor $\bd=\Idd-\bPhi^{-2}$ is equal to 1,
whatever the stress multiaxiality and without the need of a procedure bounding the damage eigenvalues to 1. 
}

{
To illustrate this property, we describe below the three particular cases of uniaxial, equi-biaxial and equi-triaxial tension loadings. 
It is shown that, at low damage in those three loading cases, one recovers the expression $\left.\tilde K=K(1-\eta d_H)\right.$ due to \cite{LDS2000}.
\begin{itemize}
\item[--]  \emph{In uniaxial tension} the damage tensor of quasi-brittle materials is classically $\bd=\textrm{diag}[d_1, 0, 0]$ \citep{LK1993,Kra1996} and $\Phi_1=\left(1-d_1\right)^{-1/2}\geq 1$, $\Phi_2=\Phi_3=1$ so that the effective bulk modulus \eqref{eq:KtildeetaDH} has for expression
\beq\label{eq:KtT}
	\tilde K =\displaystyle 
	\frac{5K(1-3 d_H)}{5-(15-8 \eta)d_H -2\eta \left(1-\sqrt{1-3 d_H}\right)}.
\eeq
In this uniaxial loading, the  maximum principal damage is $d_1=3d_H$.
\item[--]  \emph{The equi-triaxial tension case} corresponds to spherical damage tensors $\bd=d_H\,\Idd$,  $\bPhi=\Phi_H\, \Idd=(1-d_H)^{-1/2} \Idd $,
with thus $\tr \bPhi^2=\frac{1}{3}(\tr \bPhi)^2=3/(1-d_H)$, 
and so Eq. \eqref{eq:KtildeetaDH} rewrites as
\beq\label{eq:KtTT}
	\tilde K = \frac{K(1- d_H)}{1-(1-\eta) d_H}.
\eeq
In this equi-triaxial loading, the maximum principal damages are $d_i=d_H$. Note that, then, the value $\eta=1$ leads to the linear law $\left.\tilde K = K(1- d_H)\right.$ over the whole range of damage.
\item[--]  We can also consider \emph{equi-biaxial tension} for which $d_1=d_2\geq 0$, $d_3=0$, $\Phi_1=\Phi_2=(1-d_1)^{-1/2}\geq 0$, $\Phi_3=1$. The effective bulk modulus \eqref{eq:KtildeetaDH} becomes then
\beq\label{eq:KtBT}
	\tilde K = \frac{5K(2-3 d_H)}{10- (15-13 \eta )d_H  -2\eta \left(2-\sqrt{4-6 d_H}\right) }.
\eeq
In this equi-biaxial loading, the maximum principal damages are $d_1=d_2=\frac{3}{2}d_H$.
\end{itemize}
}

\begin{figure}[htbp]
\begin{center}
\includegraphics[width=70mm]{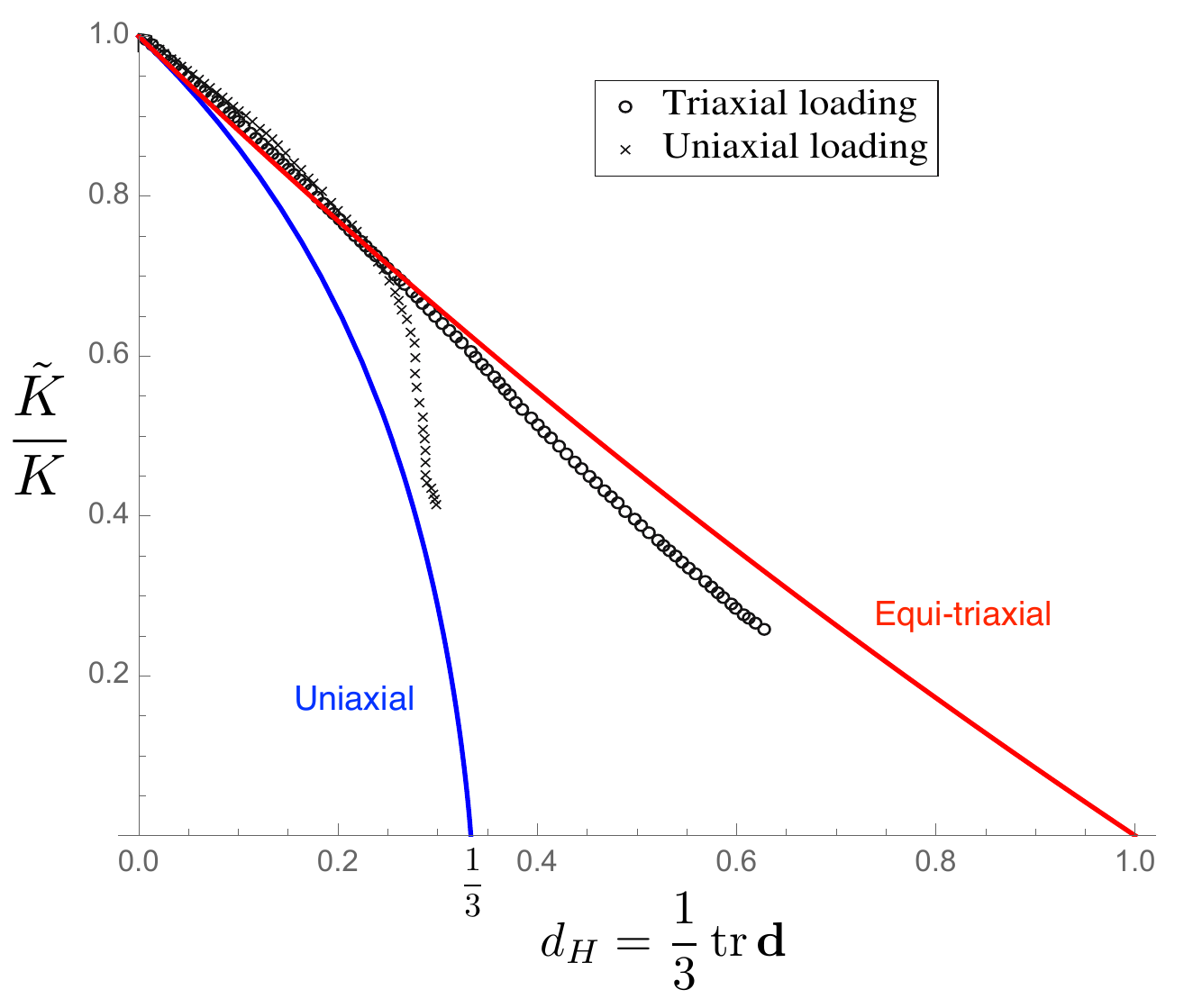}
\caption{Effective bulk modulus $\tilde K$ from \eqref{eq:KtildeetaDH} as a function of hydrostatic damage $d_H$  with $\bd=1-\bPhi^{-2}$ (for value $\eta=1.2$ representative of a micro-concrete).\label{fig:KtildeH}}   
\end{center}
\end{figure}

{
Figure \ref{fig:KtildeH} shows that, for the whole range of hydrostatic damage, the loss of bulk modulus  \eqref{eq:KtildeetaDH} with $\eta=1.2$ behave in a similar manner as the one of Fig. \ref{fig:KtildeArnaud_DH} obtained from Discrete Element computations.
}

{
As expected, the effective (damaged) bulk modulus $\tilde K$ vanishes exactly when maximum principal damage(s) are equal to 1 (solid lines in Fig. \ref{fig:Ktildemaxdi}) in all these loading cases. The first order expansions in $d_H$ (at small damage) of the three Equations \eqref{eq:KtT}, \eqref{eq:KtTT} and \eqref{eq:KtBT} gives for all three---uniaxial, equi-biaxial and equi-triaxial---loading cases
\beq
	\tilde K \approx K(1-\eta d_H)
\eeq
(dashed lines in Fig. \ref{fig:Ktildemaxdi})
{\it i.e.}  the exact expression introduced by \cite{LDS2000} in a fully phenomenological manner 
(recovering Eq. \eqref{eq:KtildedHDD} when $\eta=1.2$ is set).
}

\section{Conclusion}
\label{sec:conclusion}

Some mathematical tools such as the {harmonic product} and the {harmonic factorization into lower order tensors} have been presented. Together with the notion of \emph{mechanically accessible directions for measurements}, this has allowed us to derive, at harmonic order 4, both a crack density expansion $\Omega(\nn)$ and a {micro-mechanics based} damage framework that makes use of second-order tensorial variables only, instead of fourth-order in standard micro-mechanics based approaches. {The hydrostatic sensitivity obtained from such a second-order damage framework is shown to have the sought property of leading to the vanishing of  the effective (damaged) bulk modulus at maximum principal damage $\max d_i$ exactly equal to 1.}

\appendix

\section{{Spherical/Deviatoric harmonic decomposition}}
\label{A:phiphi}

{
Different practical expressions of the harmonic decomposition of a fourth order tensor of elasticity type exist \citep{Bac1970,Ona1984,Auffray2017}. We use here the spherical/deviatoric harmonic decomposition introduced by \cite{Auffray2017} of a fourth order tensor $\tq T$ of the elasticity type
\begin{multline}\label{eq:HarmDecomp}
\tq T = \alpha \; \Idd \otimes \Idd 
+ 2 \beta \; \tq J
+ \Idd \otimes \bc'+ \bc' \otimes \Idd
\\
+ 2\Big((\Idd \otimesbar \bb' +\bb'  \otimesbar \Idd) -\frac{2}{3}(\Idd \otimes \bb'+ \bb' \otimes \Idd)\Big)
+ \bH,
\end{multline}
where, $\td{di}$ and $\td{vo}$ being the dilatation and voigt tensors of $\bT$ and
$\td{di}'$ and $\td{vo}'$ their deviatoric parts,
\begin{align*}
\td{di}&= \tr_{12} \tq T,
& 
\td{vo} &= \tr_{13} \tq T,
\\
\alpha &= \frac{1}{9}\tr{\td{di}},
&
\beta&=\frac{1}{30}(- \tr{\td{di}} +3\tr{\td{vo}}),
\\
\td{c}' &=\frac{1}{3}\td{di}',
&
\td{b}' &=\frac{1}{7}(-2 \td{di}' +3 \td{vo}').
\end{align*}
}

{
Introducing two second symmetric second order tensors $\td y$ and $\td x$ such that
$\td y = \tq T : \td x$, the following equalities are obtained :
\begin{align*}
\td x :\tq T : \td x =
\alpha (\tr \td x)^{2} &+ 2 \beta \td{x}':\td{x}'
+2 (\tr \td x)(\bc':\td{x}')
\\
&+4 \bb' : \td{x'}^{2} + \td{x}':\bH:\td{x}',
\end{align*}
and
\begin{align*}
\td{y}&= \alpha (\tr \td{x}) \Idd + 2 \beta \td{x}' + (\bc':\td{x}') \Idd+(\tr \td{x}) \bc' 
\\
 & \qquad \qquad + 2 \left[ (\bb'.\td{x}')'+(\td{x}'.\bb')'\right] + \bH:\td{x}',
\\
\tr \td y &= 3 (\alpha \tr \td{x} + \bc' : \td{x}'),
\\
\td{y}'&= (\tr \td{x}) \bc' + 2 \beta \td{x}' 
 + 2 \left[ (\bb'.\td{x}')'+(\td{x}'.\bb')'\right] + \bH:\td{x}'.
\end{align*}
}

\section{{Harmonic decomposition of 
$\tq D$}}
\label{A:HarmDecompD}

{
From equation~\eqref{eq:Gibbsg}, the elasticity law is obtained as
\begin{equation}\label{eq:AppElastgCompl}
  \tilde {\tq S}= \frac{g(\bPhi)}{9K} \Idd \otimes \Idd+\frac{1}{2G} \tq G  ,
\end{equation}
with fourth order tensor
\begin{equation} \label{eq:AppdefG}
  \tq G = \bPhi\otimesbar \bPhi+\frac{1}{9}(\tr \bPhi^2)\, \Idd \otimes \Idd -\frac{1}{3} (\Idd \otimes \bPhi^2 +\bPhi^2 \otimes \Idd).
\end{equation}
The harmonic decomposition~\eqref{eq:HarmDecomp} of $\bPhi\otimesbar \bPhi$ reads
\begin{multline*}
\bPhi\otimesbar \bPhi = \alpha \; \Idd \otimes \Idd 
+ 2 \beta \; \tq J
+ \Idd \otimes \bc'+ \bc' \otimes \Idd
\\
+ 2\left((\Idd \otimesbar \bb' +\bb'  \otimesbar \Idd) -\frac{2}{3}(\Idd \otimes \bb'+ \bb' \otimes \Idd)\right)
+ \bH
\end{multline*}
with
\begin{align*}
\td{di}&= \bPhi^{2},
& 
\td{vo} &= \frac{1}{2}\left[ (\tr \bPhi) \bPhi + \bPhi^{2}\right],
\\
\alpha &= \frac{1}{9}\tr (\bPhi^{2}),
&
\beta&=\frac{1}{60}\left[ \tr (\bPhi^{2}) + 3 (\tr \bPhi)^{2}\right],
\\
\td{c}' &=\frac{1}{3} (\bPhi^{2})',
&
\td{b}' &=\frac{1}{14}\left[ 3 (\tr \bPhi) \bPhi'-(\bPhi^{2})'\right].
\end{align*}
As $(\bPhi\otimesbar \bPhi)^{S}=\bPhi \odot \bPhi$, we get 
\begin{equation*}
\bH = ((\bPhi\otimesbar \bPhi)^{S})_0 = (\bPhi\odot \bPhi)_0=\bPhi\ast \bPhi.
\end{equation*}
Noting that
\begin{align*}
\alpha \; \Idd \otimes \Idd &+\Idd \otimes \bc'+ \bc' \otimes \Idd \\
=&+\frac{\tr{\bPhi^{2}}}{9} \; \Idd \otimes \Idd +\frac{1}{3}\left(\Idd \otimes (\bPhi^{2})'+ (\bPhi^{2})' \otimes \Idd\right)
\\
=&-\frac{\tr{\bPhi^{2}}}{9} \; \Idd \otimes \Idd +\frac{1}{3}\left(\Idd \otimes (\bPhi^{2})+ (\bPhi^{2}) \otimes \Idd\right)
\end{align*}
equations~\eqref{eq:AppElastgCompl} and~\eqref{eq:AppdefG} lead to
\begin{equation}
 \tilde {\tq S}= \frac{1}{9K} \Idd \otimes \Idd +\frac{1}{2G} \tq J +\frac{1}{E} \bD ,
\end{equation}
with
\begin{align*}
 \frac{1}{E}\bD= &\frac{g(\bPhi)-1}{9K} \Idd \otimes \Idd
 +\frac{2 \beta -1}{2 G} \tq J
 \\
  &+ \frac{1}{G}\left(\Idd \otimesbar \bb' +\bb'  \otimesbar \Idd -\frac{2}{3}(\Idd \otimes \bb'+ \bb' \otimes \Idd)\right)
 \\
&+ \frac{1}{2 G} \bPhi \ast \bPhi.
\end{align*}
A straightforward calculation leads to the property 
\begin{align*}
\left(\tr_{12} \bD\right)' &=\pmb{0},
&
\left(\tr_{13} \bD\right)' &=\frac{7(1+\nu)}{3 } \bb'.
\end{align*}
}

\section*{References}


\end{document}